\documentclass[aps,prl,amsmath,twocolumn,superscriptaddress]{revtex4-1}
\usepackage{graphicx}
\graphicspath{ {images/} }
\usepackage{dcolumn}
\usepackage{bm}
\usepackage{color,soul}

\begin{document}

\preprint{APS/123-QED}

	\author{S. Flannigan} \affiliation{Department  of  Physics  $\&$  SUPA,  University  of  Strathclyde,  Glasgow  G4  0NG,  United  Kingdom.}
	\author{F. Damanet}\affiliation{Institut de Physique Nucléaire, Atomique et de Spectroscopie, CESAM, University of Liège, B-4000 Liège, Belgium}
	\author{A. J. Daley} \affiliation{Department  of  Physics  $\&$  SUPA,  University  of  Strathclyde,  Glasgow  G4  0NG,  United  Kingdom.}


\title{Many-body quantum state diffusion for non-Markovian dynamics in strongly interacting systems}

\date{\today}

\begin{abstract}
Capturing non-Markovian dynamics of open quantum systems is generally a challenging problem, especially for strongly-interacting many-body systems. In this work, we combine recently developed non-Markovian quantum state diffusion techniques with tensor network methods to address this challenge. As a first example, we explore a Hubbard-Holstein model with dissipative phonon modes, where this new approach allows us to quantitatively assess how correlations spread in the presence of non-Markovian dissipation in a 1D many-body system. We find regimes where correlation growth can be enhanced by these effects, offering new routes for dissipatively enhancing transport and correlation spreading, relevant for both solid state and cold atom experiments.  
\end{abstract}

\maketitle

\textit{Introduction.}
In open quantum system dynamics, it is becoming increasingly crucial to consider the effects of non-Markovian dissipation, i.e., dissipation into a spectrally-structured environment which remembers past interactions with the system~\cite{RevModPhys.89.015001}, as demonstrated in many recent quantum devices which are non-Markovian in nature~\cite{PhysRevLett.122.050501,Liu:2011vt,PhysRevX.8.011053}. 
While there has been great progress in treating these features computationally~\cite{Ishizaki:2005aa,Kato:2016aa,Strathearn:2018aa,PhysRevResearch.2.013265,PhysRevLett.126.200401,PhysRevLett.113.150403,Hartmann:2017aa}, there has so far been difficulty in generalising these methods for strongly-interacting many-body systems, even in 1D. Here, by hybridizing tensor network and non-Markovian stochastic techniques, we show how to capture the effects of non-Markovian dissipation on the generation of long-range correlations in strongly interacting one-dimensional many-body systems. As an example, we consider a damped form of the Hubbard-Holstein model, which introduces electron-phonon interactions to strongly correlated systems~\cite{Holstein:1959aa,Holstein:2000aa,doi:10.1142/1476}. We find that the growth of pairing correlations can be enhanced by going beyond the Markovian regime and that by controlling the properties of the environment we can tune the correlation spreading in the (electron) system.
Our results demonstrate the capabilities of these methods to explore dissipative many-body systems beyond the Born-Markov limit and quantitatively capture their out-of-equilibrium dynamics, as motivated by experimental advances with many-body cavity quantum electrodynamics (QED)~\cite{PhysRevX.8.011002,Kollar:2017vn,PhysRevA.87.043817} and with cold atoms immersed in reservoir gases~\cite{PhysRevLett.94.040404,PhysRevA.84.031602,PhysRevA.95.033610,PhysRevA.98.062106,PhysRevA.101.033612}.

%

\begin{figure}[t!]
{\includegraphics[width = 8cm]{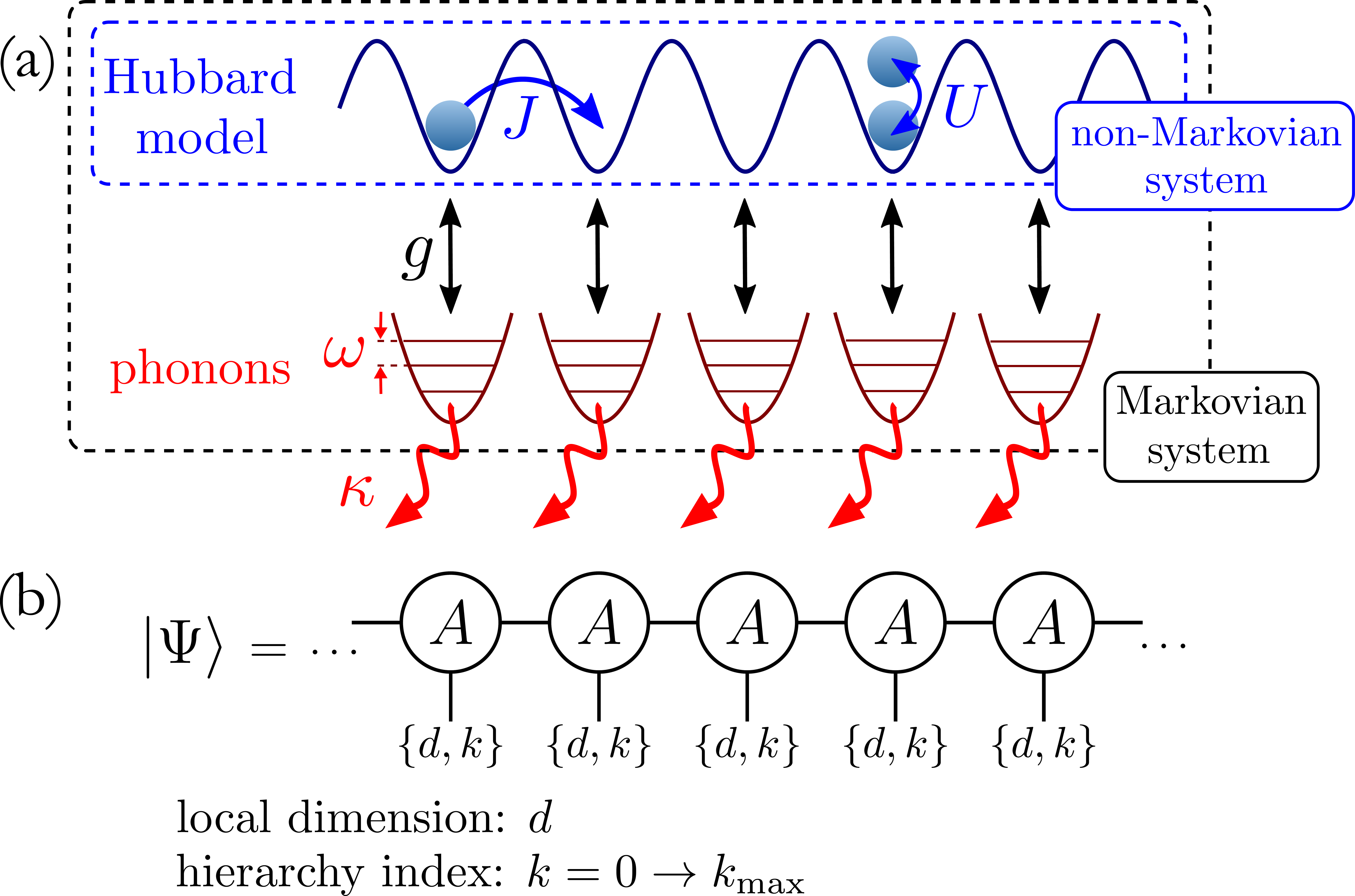}}\\ 
\caption{(a) Illustration of the Hubbard model coupled with strength $g$ to independent identical local phonon modes of frequency $\omega$ and damping rate $\kappa$. While the dissipative dynamics of the system made up of the fermions and the phonons (dashed black box) is Markovian, the one of the Hubbard system alone (dashed blue box) is generally non-Markovian. (b) Matrix product state (MPS) representation of the many-body HOPS equations [see Eq.~(\ref{HOPS_MPS})], with local dimension $d$ and hierarchy dimension $k_{\rm max} + 1$ in the usual form but now with an enlarged local dimension $d + k_{\rm max} + 1$. }
\label{NMSystem}
\end{figure}

Large separations of frequency scales in quantum optical systems coupled to their environment have made theoretical tools such as the Gorini, Kossakowski, Sudarshan, Lindblad (GKSL) 
master equation~\cite{lindblad1976,Gorini:1976aa} invaluable for quantitatively capturing many important experiments. There, the system and environment are weakly coupled and the environment is memory less, satisfying the Born-Markov approximation~\cite{Qnoise,Breur_book}. 
Reservoir engineering in recent quantum optics experiments, such as using impurities immersed in Bose-Einstein Condensates (BEC) to produce spin-boson models~\cite{PhysRevLett.94.040404,PhysRevA.84.031602,PhysRevA.95.033610,PhysRevA.98.062106,PhysRevA.101.033612} or with multi-mode cavity QED systems~\cite{PhysRevX.8.011002,Kollar:2017vn,PhysRevA.87.043817}, has made it possible to go beyond the Born-Markov regime in systems where microscopic models can still be derived from first principles. This has motivated interest in creating theoretical tools to compute dynamics in these cases. The large size of these systems makes it necessary to trace out the BEC in the former scenario and the cavity modes in the latter, which results in open quantum system descriptions that are generally non-Markovian~\cite{Caldeira:1983wy,RevModPhys.59.1,doi:10.1142/1476,PhysRevA.99.033845,PhysRevResearch.3.L032016}. Simulating these situations is particularly challenging due to the combination of strong interactions generating strongly correlated phases, the many-body system giving rise to an exponentially large Hilbert space and the non-Markovian features requiring the use of an equation of motion that is non-local in time. 

Finding the best way to deal with non-Markovian dynamics, the most natural kind of open system dynamics occurring in the solid-state from which our example originates, is an old and difficult problem, and a number of approaches have been developed over the past decades, ranging from non-Markovian master equations
~\cite{Breur_book} to non-Markovian collapse theories~\cite{PhysRevA.80.012116}, collisional models~\cite{Rybar:2012un} and stochastic Schrödinger equations~\cite{DIOSI1997569, Piilo2008PRL} (see~\cite{RevModPhys.89.015001} for a detailed review).
More recently, time-evolving matrix product operators (TEMPO)~\cite{Strathearn:2018aa,PhysRevResearch.2.013265,PhysRevLett.126.200401} or hierarchical equations of motion (HEOM)~\cite{Ishizaki:2005aa,Kato:2016aa} have shown remarkable potential for systems with a small Hilbert space, but so far have not been generalised to many-body systems.


To address this challenge here we employ the hierarchy of pure states (HOPS)~\cite{PhysRevLett.113.150403,Hartmann:2017aa}, a non-Markovian quantum state diffusion method, which we have combined with matrix product state (MPS) techniques~\cite{SCHOLLWOCK201196}. We demonstrate applications for this method by exploring dynamics in a modified Hubbard-Holstein model \cite{Holstein:1959aa,Holstein:2000aa}, where we couple strongly interacting fermions to local harmonic oscillator modes that are damped, representative of phonons that have dispersion. We show that non-Markovian dissipation can enhance the short-time dynamical growth of the pairing correlations where we find a qualitative difference compared to the Markovian, but also the phononless cases. This demonstrates that this method allows us to quantitatively simulate the dynamics of strongly correlated one-dimensional open many-body systems well into the non-Markovian and strong coupling regimes.

\textit{The dissipative Hubbard-Holstein model.}
We consider the model shown in Fig.~\ref{NMSystem}(a), with fermions in an $M$ site lattice, described by a many-body system Hamiltonian $\hat{H}_s$ where each site is coupled to a local phonon mode similar to the (Hubbard)-Holstein model~\cite{Holstein:1959aa,Holstein:2000aa}. The total Hamiltonian is given by
\begin{equation}\label{Open_problem}
\hat{H} = \hat{H}_s + \omega \sum_{n=1}^M  \hat{a}_n^{\dagger} \hat{a}_n + g\sum_{n=1}^M \Big( \hat{L}_n\hat{a}^{\dagger}_n + \hat{L}_n^{\dagger}\hat{a}_n \Big),
\end{equation}
where $\hat{a}_n^{\dagger}$ and $\hat{a}_n$ create and destroy a phonon in the $n$th mode and $\hat{L}_n$ are system operators acting on site $n$. 
We modify the usual Holstein model by going beyond the approximation of dispersionless phonons, taking a next step in better modelling realistic situations with this toy model~\cite{PhysRevLett.120.187003}. We incorporate these effects by modelling each phonon mode as a damped harmonic oscillator, such that we can write the phonon correlation function as,
\begin{equation} \label{Ph_Corr}
\alpha_{n}(t-t') = \langle \hat{a}_n(t) \hat{a}_n^{\dagger}(t') \rangle = e^{- \kappa |t-t'| - i \omega (t-t')},
\end{equation}
where $\omega$ and $\kappa$ are the phonon frequency and damping rate, respectively. 

\textit{Non-Markovian Quantum State Diffusion.}
Non-markovian dynamics arise when we trace out part of the system where we do not have a strong separation of frequency scales that satisfy the conditions for the Born-Markov approximation. In principle it is always possible to place the boundary of the system where the dynamics are Markovian. In this case, we could take the fermions and phonon modes as the \textit{system} [dashed black box in Fig.~1(a)], with the phonon damping remaining  Markovian~\cite{Daley:2014aa}.
However, in many relevant situations (such as multi-mode cavities described above), it becomes prohibitively expensive computationally to make this choice because of the large local basis. In this particular case, we find it much more convenient to trace out the phonon modes and work with an effective equation of motion for the Hubbard system only [dashed blue box in Fig.~1(a)]. For finite $\kappa$ the resulting correlation function for the phonon modes, Eq.~(\ref{Ph_Corr}), cannot be approximated as a delta function, and so we must use the non-Markovian quantum state diffusion (NMQSD) equation for the dynamics of the reduced system $|\psi(t) \rangle$~\cite{DIOSI1997569, RevModPhys.89.015001},
 \begin{equation} \label{NMQSD_eq}
 \begin{split}
\partial_t|{\psi}(t) \rangle = & -i \hat{H}_s |\psi(t) \rangle + g\sum_{n=1}^M \hat{L}_n z^*_n(t) |\psi(t) \rangle \\
& - g\sum_{n=1}^M \hat{L}_n^{\dagger} \int_0^t ds \alpha_n^*(t-s) \frac{\delta  |\psi(t) \rangle}{\delta z_n^*(s)},
\end{split}
\end{equation}
where we have introduced a set of stochastic \textit{colored} noise terms $z^*_n(t)$ which upon taking an ensemble average give the correlation function $\mathcal{E}[z_n(t) z^*_{n'}(t')] = \delta_{n,n'} \alpha_n(t-t')$. 

\textit{The HOPS + MPS algorithm.}
The insight which lead to the HOPS algorithm~\cite{PhysRevLett.113.150403,Hartmann:2017aa} is to introduce a set of auxiliary states which absorb the numerically intractable functional derivatives  $\delta/\delta z_n^*(s)$,
\begin{equation} 
|\psi^{(1,n)}(t)\rangle = D_n(t) |\psi(t) \rangle \equiv \int_0^t ds \alpha_n^*(t-s) \frac{\delta  |\psi(t) \rangle}{\delta z_n^*(s)}.
\end{equation}
Deriving an equation of motion for this auxiliary state requires the introduction of further auxiliary states defined through $|\psi^{(k,n)}(t)\rangle = [D_n(t)]^k |\psi(t) \rangle$ which give rise to a hierarchical set of equations. In order to write this hierarchy, we find it convenient to include the hierarchy index into the basis states and write a total state for the combined system and auxiliary Hilbert space, 
\begin{equation}
|{\Psi} (t) \rangle = \sum_{\vec{\mathbf{k}}} C_{\vec{\mathbf{k}}}(t) |\psi^{(\vec{\mathbf{k}})}(t)\rangle \otimes |\vec{\mathbf{k}}\rangle, 
\end{equation}
where the $C_{\vec{\mathbf{k}}}(t)$ are time-dependent complex numbers and $|\vec{\mathbf{k}}\rangle = | k_1,k_2,\cdots,k_M \rangle = |k_1\rangle \otimes |k_2 \rangle \otimes \cdots \otimes |k_M \rangle$ with each of the $k_n$ running from $0,1,\cdots,\infty$, as we have a hierarchy index for each of the $M$ phonon environment modes. Each hierarchy index is represented as an independent boson mode, see the supplementary material for details. Note that the $|\psi^{(0)}(t)\rangle \otimes |0\rangle =  |\psi(t)\rangle$ is the physical system state. This allows us to write the equation of motion for the total state as,
\begin{equation}\label{HOPS_MPS}
\begin{split}
\partial_t|{\Psi} (t) \rangle = &-i \hat{H}_s|\Psi(t) \rangle +  \sum_{n=1}^M \Big( \tilde{z}^*_n(t) g \hat{L}_n - \left( \kappa + i \omega \right)  \hat{K}_{n}    \\
&  + g\hat{L}_n \otimes \hat{K}_{n} \hat{b}_n^{\dagger} - g \left( \hat{L}_n^{\dagger} - \langle \hat{L}_n^{\dagger} \rangle_t \right) \otimes \hat{b}_n \Big) |\Psi(t) \rangle.
\end{split}
\end{equation}
Note that we time-dependently modify the colored noise according to $\tilde{z}^*_n(t) = z^*_n(t) + g\int_0^t ds \alpha^*_n(t - s) \langle \hat{L}_n^{\dagger} \rangle_s$ with $\langle \hat{L}^{\dagger}_n \rangle_s = \langle {\psi}^{(0)}(s) | \hat{L}^{\dagger}_n | {\psi}^{(0)}(s) \rangle$ thus explicitly taking into account previous states of the system. Note that one has to consider sufficiently small time steps in the numerical resolution of the equation so that the time-dependent terms in Eq.~\ref{HOPS_MPS} can be approximated as constant in time. In this way the non-linear terms $\tilde{z}^*_n(t)$ and $\langle \hat{L}_n^{\dagger} \rangle_t$ are calculated using the state before the time increment.

In the above equation we have introduced the bare operators (ommitting the index $n$) $\hat{b}^{\dagger} |k\rangle  = |k+1 \rangle$, $\hat{b} |k \rangle  = |k-1 \rangle$ (see Ref.~\cite{arxiv_bare,PhysRevA.60.4083}) and $\hat{K} = \sum_k k |k\rangle \langle k |$. We initialise the hierarchy with $C_\mathbf{0}(0) = 1$ and $C_{|\mathbf{k}|>0}(0) = 0$ and in order to extract observables we use the (normalized) physical system state $O(t) = \langle {\psi}^{(0)}(t) | \hat{O} | {\psi}^{(0)}(t) \rangle$ which we must average over many trajectories with different realisations of the random numbers $z_n^*(t)$, similar to conventional QSD equations~\cite{RevModPhys.89.015001,Qnoise,Daley:2014aa}.

Formally the hierarchy depth is infinite, but the populations of the auxiliary states typically decrease with the hierarchy indices $k_n$, which makes it possible in practice to truncate each hierarchy to some index $k_\mathrm{max}$ (chosen such that the results have converged to a given precision) to render the problem numerically feasible.
In general, the stronger the violation of the Born-Markov approximations the larger the number of auxiliary states we must retain. Note that this hierarchy truncation still results in an exponential number of equations: if each hierarchy index can run from $0,1,\cdots,k_{\rm max} $ then in total we have $({k_{\rm max}+1})^M$ auxiliary states. This motivates the incorporation of MPS techniques which allow us to time-evolve many-body states of one-dimensional Hamiltonians without explicitly working with the full Hilbert space~\cite{SCHOLLWOCK201196}. As each hierarchy only couples locally with a system operator of site $n$ this allows us to efficiently write this problem as an MPS simply with an enlarged local Hilbert space consisting of the physical local dimension of the system, but now also an effective local dimension for the auxiliary state of that effective environment mode [see Fig.~\ref{NMSystem}(b)] modelled as a boson Hilbert space. This is particularly convenient as we can then apply standard MPS techniques for time-evolution~\cite{Paeckel:2019aa}. This does result in an MPS with a large local dimension but in the following sections we show that it can be used to make important quantitative predictions with practical numerical values for the size of the hierarchy dimension $k_{\rm max} $ and also the bond dimension of the MPS $D$ (see the supplemental material for a detailed error analysis). Note finally that providing $k_{\rm max} $ and $D$ are large enough, Eq.~(\ref{HOPS_MPS}) numerically converges to the \textit{exact} dynamics of the system (as well as of the environment via monitoring of the noises as we will discuss below), as it does not directly rely on any approximation (neither Born nor Markov).

\textit{Benchmarking.}
We first consider the out-of-equilibrium dynamics of a Holstein model~\cite{Holstein:1959aa,Holstein:2000aa}. We use,
\begin{equation}\label{Hol_Ham}
\hat{H}_s = -J \sum_n \left( \hat{c}^{\dagger}_n \hat{c}_{n+1} + \hat{c}^{\dagger}_{n+1} \hat{c}_n \right),
\end{equation}
as the system Hamiltonian in Eq.~(\ref{HOPS_MPS}), where $J$ describes the tunnelling of the (spinless) electrons. Additionally, we use the number operator as our system-environment coupling operators $\hat{L}_n = \hat{n}_n = \hat{c}^{\dagger}_n \hat{c}_{n}$, and as mentioned we include dissipation on the phonons yielding the damped correlation functions, Eq.~(\ref{Ph_Corr}).

\begin{figure}[t!]
\includegraphics[width=8.5cm]{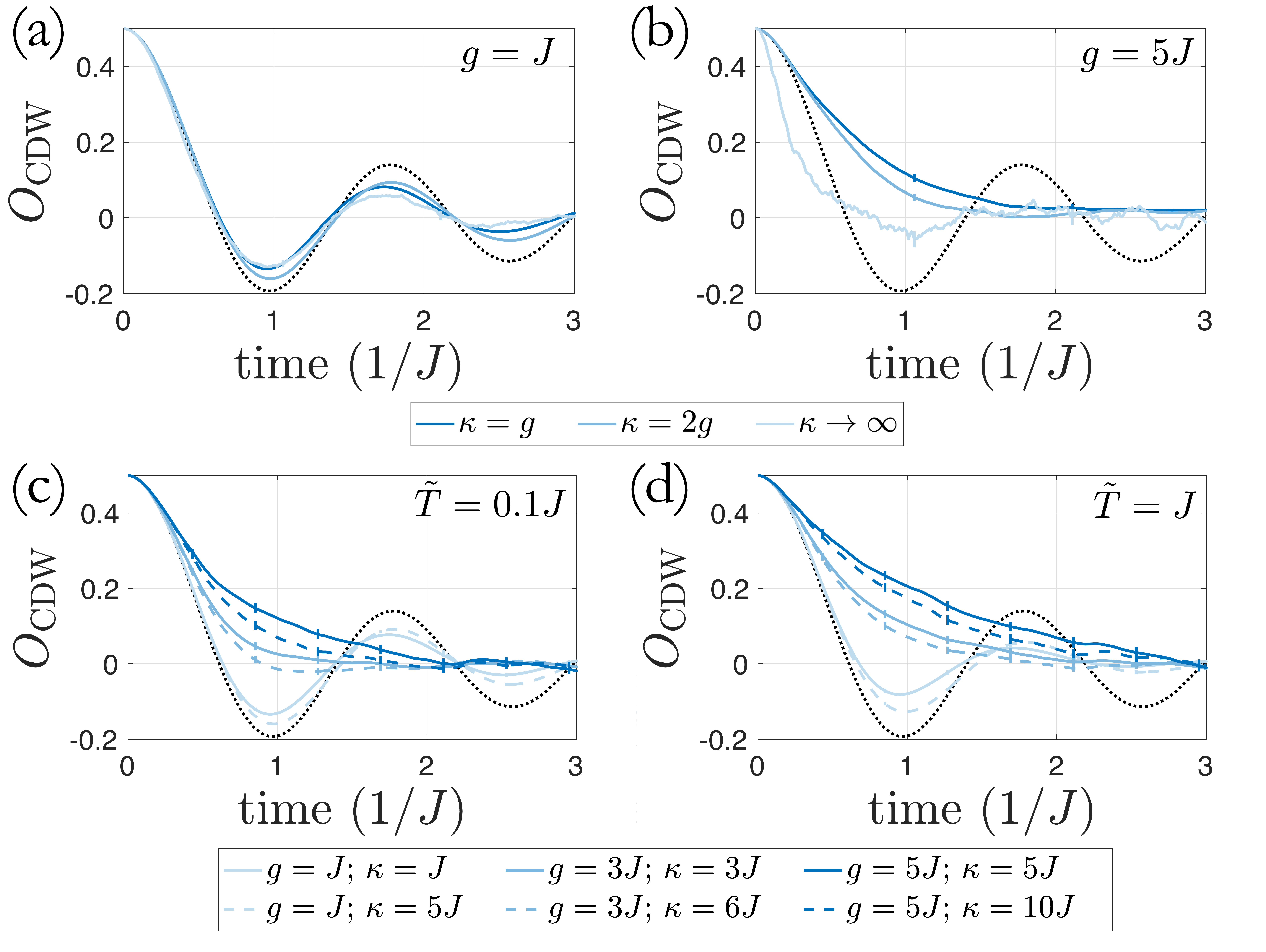}
\centering
\caption{Dynamics in the dissipative Holstein model [Eq.~(\ref{Open_problem}) with the system Hamiltonian~(\ref{Hol_Ham}), $\hat{L}_n = \hat{n}_n$ and the environment correlation functions~(\ref{Ph_Corr})] upon beginning in an initial state $|1,0,1,0,\cdots \rangle$. (a-b) The evolution of the CDW correlations $O_{\rm CDW} (t) = (1/M) \sum_n (-1)^n \langle \hat{n}_n(t) \rangle$ for different coupling strengths $g$ ($g=0$ in black dotted) and phonon dispersion rates $\kappa = [ g,2g,\infty ]$ (dark to light blue). The Born-Markov limit ($\kappa \rightarrow \infty$) was calculated using a conventional quantum trajectory method~\cite{Daley:2014aa}) (c-d) Finite temperature analysis for different coupling strengths $g = [ 1,3,5 ] J$ (light to dark blue, $g=0$ in black dotted) and phonon dispersion rates $\kappa$.  See Ref.~\cite{Hartmann:2017aa} on how to adapt the algorithm for finite temperature environments.
In all cases we average the observables over $N_{\rm traj}=100$ trajectories and use $\omega = J$ and $M=20$ lattice sites. For our hybridized HOPS+MPS algorithm, we use the numerical parameters $k_{\rm max}=8$, $D=128$ and $Jdt=0.01$.}
\label{Fig_CDW}
\end{figure}

We begin with the initial state $|1,0,1,0,\cdots \rangle$ and time-dependently calculate a charge density wave (CDW) correlations $O_{\rm CDW} (t) = (1/M)\sum_n (-1)^n \langle \hat{n}_n(t) \rangle$. We plot this in Fig.~\ref{Fig_CDW}(a-b) for different coupling strengths $g$ and phonon dispersions $\kappa$. Comparing to the results obtained in Ref.~\cite{PhysRevB.101.035134}, which analyzes this system in the limit of dispersionless phonons ($\kappa \rightarrow 0$), we find the same qualitative behaviour, where for $g=J$ the dynamics are similar to the closed system ($g=0$) case where there are oscillations but the CDW \emph{melts} into a homogeneous steady state. Increasing the coupling strength to $g=5J$ we can see that the CDW melting is slowed for short times and the oscillations become completely damped.  

\textit{Born-Markov limit.} We also compare our results to that of a conventional quantum state diffusion (QSD) equation valid in the Born-Markov limit~\cite{Daley:2014aa}. This is achieved by setting $k_{\rm max}=1$ and $\alpha_{n}(\tau) = \delta(\tau)$ (see Ref.~\cite{PhysRevLett.113.150403} and the supplemental material) which physically corresponds to the approximation that the phonon dispersion $\kappa$ goes to infinity.
From Fig.~\ref{Fig_CDW} we see that for strong coupling ($g = 5J$) this model completely fails to predict the suppression of the CDW correlations at short times. 

\textit{Finite temperature.} Within the framework of HOPS it is also possible to efficiently include finite temperature effects of the environment (see Ref.~\cite{Hartmann:2017aa}). In Fig.~\ref{Fig_CDW}(c-d) we plot the dependence on the CDW correlations upon increasing the initial temperature of the environment modes. We see that the suppression of the CDW melting is enhanced for increasing temperatures which is due to a non-zero population of phonons in the initial state, allowing for a greater effect on the short time dynamics. Including finite temperature effects in the Born-Markov QSD simply increases the effective system-environment coupling strength (see the supplemental material) which as seen from (a-b) predicts an increased decay of the CDW. Increasing the temperature of the phonon modes in this model therefore results in further deviations from the Born-Markov regime, in contrast to the more common cases where larger temperatures suppress non-Markovian features~\cite{RevModPhys.89.015001,doi:10.1142/1476}.

\textit{Correlation spreading.}
We move on and consider the Hubbard-Holstein model describing two-species fermions coupled to phonon modes and now with an onsite interaction $U$. Explicitly our system Hamiltonian is given by
\begin{equation}\label{Hubb_hol}
\hat{H}_s = -J \sum_{n,\sigma} \left( \hat{c}^{\dagger}_{n,\sigma} \hat{c}_{n+1,\sigma} + \hat{c}^{\dagger}_{n+1,\sigma} \hat{c}_{n,\sigma} \right) + U \sum_n \hat{n}_{n,\uparrow} \hat{n}_{n,\downarrow},
\end{equation}
where $\hat{n}_{n,\sigma} = \hat{c}^{\dagger}_{n,\sigma} \hat{c}_{n,\sigma}$ and our system-environment coupling operators are $\hat{L}_n = \hat{n}_{n,\uparrow} + \hat{n}_{n,\downarrow}$. As earlier, we go beyond the usual case and include phonon dissipation.

\begin{figure}[t!]
\includegraphics[width=8.5cm]{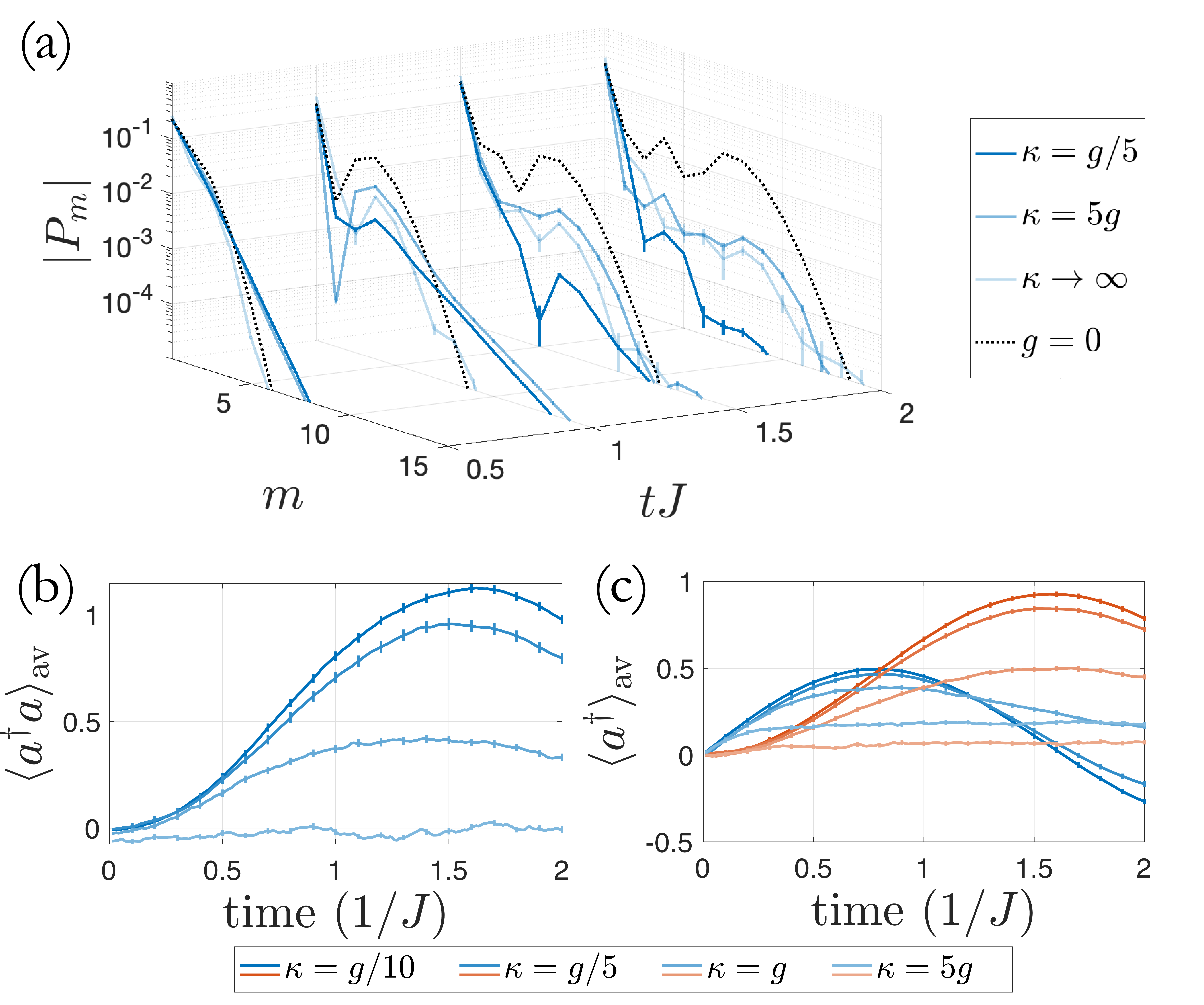}
\centering
\caption{Dynamics in the dissipative Hubbard-Holstein model [Eq.~(\ref{Open_problem}) with the system Hamiltonian~(\ref{Hubb_hol}), $\hat{L}_n = \hat{n}_{n,\uparrow} + \hat{n}_{n,\downarrow}$ and the environment correlation functions~(\ref{Ph_Corr})], upon beginning in an initial CDW state $|\uparrow,\downarrow,\uparrow,\downarrow,\cdots \rangle$. (a) The pair correlation function [Eq.~(\ref{Corr_equ})], where we use $g=J$ ($g=0$ in black) and compare different phonon dispersion rates $\kappa = [ g/5,5g,\infty ]$ (dark to light blue). (b) Dynamics of the average phonon mode occupation, $\langle a^{\dagger} a \rangle_{\rm av} = (1/M) \sum_n \langle a^{\dagger}_n a_{n} \rangle = (1/M) \sum_n \left( |\tilde{z}_n^*(t)|^2 - 1 \right) $. (c) The real (blue) and imaginary (red) part of the phonon coherences $\langle a^{\dagger} \rangle_{\rm av} = (1/M) \sum_n \langle a^{\dagger}_n \rangle = (1/M) \sum_n \tilde{z}_n^*(t)$.
For our hybridized HOPS+MPS algorithm, we use the parameters $k_{max}=6$, $D=300$ and $Jdt=0.01$ where we also have incorporated conserved quantum numbers into the MPS algorithm~\cite{itensor}. We average the observables over $N_{\rm traj}=100$  trajectories. In all cases we use $U=J$, $\omega = 2 J$ and $M=50$ sites.}
\label{Fig_HuHolCorr}
\end{figure}

In Fig.~\ref{Fig_HuHolCorr} we begin in the initial product state $|\uparrow,\downarrow,\uparrow,\downarrow,\cdots \rangle$ and in (a) we analyze the fermionic pairing correlation functions,
\begin{equation} \label{Corr_equ}
P_m = \frac{1}{M-m} \sum_{\tilde{m}} \langle \hat{c}^{\dagger}_{\tilde{m},\uparrow} \hat{c}^{\dagger}_{\tilde{m},\downarrow}  \hat{c}_{{\tilde{m}+m},\downarrow} \hat{c}_{{\tilde{m}+m},\uparrow} \rangle.
\end{equation}
For the case where there is no coupling to the phonons $g=0$ we observe a peak in these correlations which spreads out in time, and beyond this the correlations decay exponentially which is the usual light cone spreading of correlations~\cite{Lieb1972,PhysRevLett.97.050401,PhysRevB.79.155104}. Including coupling to the phonon modes with $g=J$ we see similar behaviour, although the dissipation damps the amplitude of this peak in time, gradually suppressing correlations in the steady state. For finite $\kappa$ (i.e., non-Markovian environment behaviour), we see a strong enhancement of the correlation length beyond the light cone at short times ($tJ\sim 0.5,1$) which is qualitatively different to the case of purely Markovian dissipation ($\kappa \rightarrow \infty$) where the correlation length is unaffected. 

These features of the non-Markovian dynamics can be understood by realising that the coupling to the phonons \textit{dresses} the electrons~\cite{PhysRevB.86.045110}, modifying the quasi-particle excitations and shifting the effective fermion-fermion interaction strength $U_{\rm eff} \rightarrow U - 2g^2/\omega$. Here there is a competition between a generated effective attractive interaction and then the dynamical generation of phonons in the environment, the presence of which can strongly suppress dynamics and correlation growth resulting in CDW order~\cite{PhysRevB.86.045110}. We can see the competition of these effects in Fig.~\ref{Fig_HuHolCorr}(a), which are made even more clear by analysing the phonon mode observables in Fig.~\ref{Fig_HuHolCorr}(b-c) which we can directly calculate from HOPS using the time-dependent colored noise term $\tilde{z}_n^*(t)$ in Eq.~(\ref{HOPS_MPS}) (see the supplemental material). We see that initially the phonon population is low and so the effective interaction between fermions dominates, enhancing the growth of pairing correlations, before the dynamical generation of phonons begins to dominate, suppressing correlations at later times, which for smaller $\kappa$ is larger due to an increased phonon population. 

\textit{Discussion and outlook.}
Our combination of the HOPS algorithm with MPS techniques opens up the ability to explore a wide range of new and interesting regimes that were previously only possible to simulate qualitatively and/or through invoking some strong approximations. By considering the dispersive Hubbard-Holstein model we demonstrated that we can simulate the exact dynamics of open many-body systems well into the non-Markovian and strong coupling regimes and we are able to quantitatively analyze the dynamical properties of long-range correlation functions. In particular, we found strong qualitative differences in the dynamics of fermionic pairing correlations between the non-Markovian and Markovian cases.
This work can be generalised to describe microscopic dynamics in a  range of experimental settings, such as impurities immersed in BECs~\cite{PhysRevLett.94.040404,PhysRevA.84.031602,PhysRevA.95.033610,PhysRevA.98.062106,PhysRevA.101.033612} or atoms in multi-mode cavities~\cite{PhysRevX.8.011002,Kollar:2017vn,PhysRevA.87.043817}.

Other non-Markovian techniques could be adapted in order to probe the features investigated in our work, for example TEMPO~\cite{Strathearn:2018aa,PhysRevResearch.2.013265,PhysRevLett.126.200401} or HEOM~\cite{Ishizaki:2005aa,Kato:2016aa}. It may similarly be possible to combine these methods with MPS, as we have done here with HOPS. The combination is particularly amenable to our case as it involves the evolution of a single 1D matrix product state to capture the strongly interacting open system.  
Alternatively, explicitly retaining the phonon basis states would result in an equivalent simulation with Markovian dissipation, allowing for the solution within a standard Born-Markov QSD~\cite{RevModPhys.89.015001,Daley:2014aa}. However, we find that the number of phonon basis states required (the local dimension of the MPS) is generally larger than that required for the present HOPS algorithm. In addition, HOPS has two main additional advantages. Firstly it can simulate phonon modes initially at finite temperatures and then track the induced dynamics in real time as demonstrated here, whereas explicitly retaining the basis states in this case would further increase the complexity. But secondly, improvements and extensions to the MPS representation can be immediately implemented~\cite{10.21468/SciPostPhys.10.3.058,Stolpp:2021vc,PhysRevLett.80.2661,PhysRevB.101.035134,arxiv_mps_hops}, allowing us to generalise this approach and approximate the dynamics induced by, up to reasonable timescales, environments that have algebraically decaying correlations~\cite{PhysRevLett.113.150403,Hartmann:2017aa} such as those that arise from power law spectral densities~\cite{Caldeira:1983wy,RevModPhys.59.1,RevModPhys.89.015001}.

\textit{Acknowledgements.} We thank Walter Strunz, Valentin Link, Richard Hartmann, Adrian Kantian, Sebastian Paeckel and Peter Kirton for helpful discussions.
Work at the University of Strathclyde was supported by the EPSRC Programme Grant DesOEQ (EP/P009565/1), by AFOSR grant number FA9550-18-1-0064, and by the European Union’s Horizon 2020 research and innovation program under grant agreement No.~817482 PASQuanS. F.D.\ acknowledges the Belgian F.R.S.-FNRS for financial support. Computational resources have been provided by the Consortium des Équipements de Calcul Intensif (CÉCI), funded by the Fonds de la Recherche Scientifique de Belgique (F.R.S.-FNRS) under Grant No. 2.5020.11 and by the Walloon Region.

All data underpinning this publication are openly available from the University of Strathclyde KnowledgeBase at https://doi.org/10.15129/92dd009b-00c5-4e42-ae13-38c6b21df9fd

\bibliography{HOPSTex.bib}

\newpage

\appendix

\section{Supplemental material}
 
This supplementary material gathers numerical details and additional information about the results presented in the main text. In Sec.~I we provide more details on HOPS and how it is used in practice. In Sec.~II, we present a numerical benchmarking of the hybridized HOPS + MPS algorithm. In Sec.~III, we show how we generate the stochastic processes appearing in the HOPS Eqs.~(3) and (6) of the main text. In Sec.~IV, we present a simple Markovian master equation for the Hubbard system only. Finally, in Sec.~V, we discuss how to extract the time-dependent dynamics of the environment from the algorithm.

\section{Hierarchy of Pure States}

The hierarchy of pure states (HOPS)~\cite{PhysRevLett.113.150403,Hartmann:2017aa} is a newly developed numerical method for evaluating the non-Markovian quantum state diffusion equation derived nearly 20 years prior~\cite{Strunz:1996uk,DIOSI1997569}. The insight is to extend the computational basis by introducing a hierarchy of auxiliary states, which act to facilitate the \textit{memory} effects characteristic of non-Markovian environments, thus deriving a numerically exact equation of motion without needing to invoke weak-coupling approximations or relying on large separations of timescales between the system and environment. The method is thus very general and can in principle be applied to a wide range of problems. While the initial version of this approach is only valid for environments that consist of non-interacting bosons, however, an extension to these methods for fermionic environments has also been derived~\cite{Suess:2015aa}. 

Successes of the method have already been demonstrated for the spin-boson model with a two-level atom coupled to an environment that has a power law spectral density at zero~\cite{PhysRevLett.113.150403} and finite temperatures~\cite{Hartmann:2017aa}. The method is less computationally expensive when applying it to model couplings to environments that have Lorentzian spectral densities, as considered in this work, strongly suggesting that it can have further great success upon modelling atom-only dynamics in cavity QED systems~\cite{PhysRevResearch.3.L032016,PhysRevA.99.033845}.

\subsection{Example: One Site, One Environment}

For clarity, we explicitly write down the HOPS equations first for the case where we have a single site coupled to a single damped phonon environment (i.e., case $M = 1$ in the main text). This case is particularly simple as our total hierarchy state can be modelled with a single bosonic mode, $|\vec{\mathbf{k}}\rangle = |k\rangle$.  Let us also assume that the system is a single site that can either have no particles or a single spinless fermion such that the system Hilbert space is of local dimension $d = 2$. We can write the state then in the form
\begin{equation}
|{\Psi} (t) \rangle = \sum_k  \begin{pmatrix}
 A_k(t)\\B_k(t)
\end{pmatrix} \otimes |k \rangle 
\end{equation}
where we have introduced two sets of complex numbers, $ A_k(t)$ and $B_k(t)$ to represent an arbitrary system state.

For concreteness we will also choose $k_{\rm max} = 2$, such that the hierarchy state has a Hilbert space of size $k_{\rm max} + 1 = 3$. We can then write the initial state, which for the HOPS algorithm is always initialised with only the $k=0$ auxiliary state, as
\begin{equation} \label{in_state}
|{\Psi} (0) \rangle = \begin{pmatrix}
 A_0(0)\\B_0(0)
\end{pmatrix} \otimes \begin{pmatrix}
1 \\ 0 \\ 0
\end{pmatrix}.
\end{equation}
We can see that this is simply a two-level system coupled to a single boson mode. 

As in the main text, we consider a total Hamiltonian of the form
\begin{equation}
    \hat{H} = \hat{H}_s + \omega \hat{a}^\dagger \hat{a} + g\left(\hat{L} \hat{a}^\dagger + \hat{L}^\dagger \hat{a} \right),
\end{equation}
where $\hat{L}$ acts on the system and $\hat{a}$ destroys an excitation in the physical picture of the environment. According to Eq.~(6) of the main text, the time evolution of the state $|{\Psi} (t) \rangle$ is given by,
\begin{equation}
\begin{split}\label{teeq}
\partial_t|{\Psi} (t) =& \left(-i \hat{H}_s + \tilde{z}^*(t) g \hat{L} - \left( \kappa + i \omega \right)  \hat{K} \right. \\
& \left. + g\hat{L} \otimes \hat{K} \hat{b}^{\dagger} - g\left( \hat{L}^{\dagger} - \langle \hat{L}^{\dagger} \rangle_t \right) \otimes \hat{b}, \right)|{\Psi} (t),
\end{split}
\end{equation}
where the colored noise is defined by $\tilde{z}^*(t) = z^*(t) + g\int_0^t ds \alpha^*(t - s) \langle \hat{L}^{\dagger} \rangle_s$ with $\langle \hat{L}^{\dagger} \rangle_s = \langle {\psi}^{(0)}(s) | \hat{L}^{\dagger} | {\psi}^{(0)}(s) \rangle$ and $\mathcal{E}[z(t) z^*(t')] = \alpha(t-t')$, where $\alpha(t-t') = \langle \hat{a}(t) \hat{a}^{\dagger}(t') \rangle = e^{- \kappa |t-t'| - i \omega (t-t')}$ is the environment correlation function. The operators $\hat{b}$, $\hat{b}^\dagger$ and $\hat{K}$ act on the hierarchy Hilbert space. In the basis $\{ |0\rangle, |1\rangle, |2\rangle\}$, they have the matrix representations
\begin{equation}\label{ops}
\begin{split}
\hat{K} = & \begin{pmatrix}
0~~0~~0 \\ 0~~1~~0 \\ 0~~0~~2
\end{pmatrix}, \\
\\
\hat{b} = & \begin{pmatrix}
0~~1~~0 \\ 0~~0~~1 \\ 0~~0~~0
\end{pmatrix}, \\
\\
\hat{b}^{\dagger} = & \begin{pmatrix}
0~~0~~0 \\ 1~~0~~0 \\ 0~~1~~0
\end{pmatrix}.
\end{split}
\end{equation}

While initially [Eq.~(\ref{in_state})] the system and hierarchy auxiliary states are uncorrelated, we can see from Eq.~(\ref{teeq})] that the terms $ \hat{L} \otimes \hat{K} \hat{b}^{\dagger}$ and $\hat{L}^{\dagger}  \otimes \hat{b}$ can (and generally do) generate correlations and entanglement between them. Note that, these coupling operators are dissipative, decreasing and increasing the relative norm between the different states in the hierarchy $ P_k(t) = \langle \psi^{({{k}})}(t) |\psi^{({{k}})}(t)\rangle = |A_k(t)|^2 + |B_k(t)|^2$, but can still generate entanglement. 

After the application of each numerical time-step, 
we rescale all states $|\psi^{({{k}})}(t) \rangle$ by the square root of the norm of the physical system state $|\psi^{({{0}})}(t) \rangle = \langle 0|  {\Psi} (t) \rangle$ by setting $|\psi^{({{k}})}(t)\rangle \rightarrow \frac{1}{\sqrt{P_0(t)}}  |\psi^{({{k}})}(t)\rangle$. In doing so, we enforce the normalization condition that $P_0(t) = 1$ during time evolution while keeping the relative norms $P_{k}(t)/P_{k'}(t)$ of each auxiliary state invariant. Note that we calculate the expectation values of the observables at each time-step using the normalized physical system state via
\begin{equation}
O(t) =  \frac{1}{P_0(t)} \langle \psi^{({{0}})}(t)| \hat{O} |\psi^{({{0}})}(t)\rangle,
\end{equation}
and that we must average them over many different realisations (trajectories) of the random numbers $z^*(t)$.

\subsection{Example: Two Sites, Two Environments}

Additionally, we consider how to generalise HOPS to larger systems and more environment modes by considering the case of a two-site system and two independent damped phonon environments, where as in the main text each environment couples to a single different system site with an interaction Hamiltonian, 
\begin{equation}
\hat{H}_{\rm Int} = g\sum_{n=1}^M \Big( \hat{L}_n\hat{a}^{\dagger}_n + \hat{L}_n^{\dagger}\hat{a}_n \Big).
\end{equation}

We can represent our total system-hierarchy state as,
\begin{equation}
\begin{split}
|{\Psi} (t) \rangle = \sum_{k_1,k_2}  \begin{pmatrix}
 A_{k_1,k_2} (t)\\B_{k_1,k_2} (t)\\C_{k_1,k_2} (t)\\D_{k_1,k_2} (t)
\end{pmatrix} \otimes |k_1 \rangle \otimes |k_2 \rangle,
\\
\\
\end{split}
\end{equation}
where we have introduced now four sets of complex numbers, $ A_{k_1,k_2} (t)$, $B_{k_1,k_2} (t)$,$C_{k_1,k_2} (t)$ and $D_{k_1,k_2} (t)$ to represent an arbitrary system state which now has a Hilbert space of $d^2 = 4$. For the case where $k_{\rm max} = 2$ (for both hierarchies, but note that these do not necessarily have to have the same dimension) we initialise the state as,
\begin{equation}
|{\Psi} (t) \rangle = \sum_{k_1,k_2}  \begin{pmatrix}
 A_{k_1,k_2} (t)\\B_{k_1,k_2} (t)\\C_{k_1,k_2} (t)\\D_{k_1,k_2} (t)
\end{pmatrix} \otimes \begin{pmatrix}
1 \\ 0 \\ 0
\end{pmatrix} \otimes \begin{pmatrix}
1 \\ 0 \\ 0
\end{pmatrix}.
\end{equation}
This time, our evolution equation for $|{\Psi} (t) \rangle$ contains two types of hierarchy operators acting on different sectors of the Hilbert space: $\hat{K}_1$, $\hat{b}_1$ and $\hat{b}^{\dagger}_1$, acting on the hierarchy dimension which models the first environment and $\hat{K}_2$, $\hat{b}_2$ and $\hat{b}^{\dagger}_2$ which act on the second hierarchy dimension responsible for modelling the effects of the second environment. Both sets of matrix representations are defined as in Eq.~(\ref{ops}).

\subsection{HOPS $+$ MPS for $M$ sites and $M$ Environments}

It can be seen that upon further increasing the size of the system and the number of environment modes that the Hilbert space grows exponentially thus motivating the incorporation of MPS techniques. Representing a 1D many-body system efficiently with these techniques is now quite standard, but it remains to be investigated on the best way to incorporate the hierarchy basis states. As a first demonstration here we simply incorporate each hierarchy dimension into a separate but different local dimension of the MPS. Due to the physical system that we are considering with an interaction Hamiltonian that couples an environment locally to a particular lattice site, this is a convenient representation.

Explicitly, the local dimension of the $m$th site is extended to include the basis states $| k_m \rangle $,
\begin{widetext}
\begin{equation}
\begin{split}
| \Psi(t) \rangle & = \sum_{d_1,d_2,\cdots, d_M} \sum_{k_1,k_2,\cdots,k_M} C_{(d_1,d_2,\cdots, d_M),(k_1,k_2,\cdots,k_M)} |d_1,d_2,\cdots, d_M \rangle |k_1,k_2,\cdots,k_M \rangle \\
& = \sum_{d_1,k_1,d_2,k_2,\cdots, d_M,k_M} C_{d_1,k_1,d_2,k_2,\cdots, d_M,k_M} | d_1,k_1 \rangle | d_2,k_2 \rangle,\cdots,| d_M,k_M \rangle.
\end{split}
\end{equation}
The coefficients $C_{d_1,k_1,d_2,k_2,\cdots, d_M,k_M}$ can then decomposed into a product of matrices~\cite{SCHOLLWOCK201196},
\begin{equation}
| \Psi(t) \rangle = \sum_{d_1,k_1,d_2,k_2,\cdots, d_M,k_M}A_{(d_1\times k_1)}^{1,D_1} A_{(d_2\times k_2)}^{D_1,D_2} \cdots A_{(d_M \times k_M)}^{D_{M-1},1}  | d_1,k_1 \rangle | d_2,k_2 \rangle,\cdots,| d_M,k_M \rangle.
\end{equation}
\end{widetext}
This is represented graphically in Fig.~1(b) in the main text. This representation, while most likely not being the most optimal, is particularly convenient as we can apply standard MPS techniques for time-evolution~\cite{Paeckel:2019aa} without modification, in our case we used the time-dependent variational principle (TDVP)~\cite{PhysRevLett.107.070601}.
 
\section{Numerical Benchmarking}
 
In this section we quantitatively benchmark the errors in the algorithm introduced by limiting the numerical precision. In all cases presented here, we calculate the errors induced in the Holstein model on the off-diagonal correlation functions,
 \begin{equation}
\varepsilon = \frac{1}{M^2} \sum_{n,m} \left| \langle c^{\dagger}_n c_m \rangle - \langle c^{\dagger}_n c_m \rangle_{\rm exact}  \right|,
\end{equation}
where by \textit{exact}, we mean the results predicted upon taking the limit that the bond dimension $D$ and the hierarchy depth $k_{\rm max}$ go to infinity. 

\begin{figure}[t!]
\includegraphics[width=8cm]{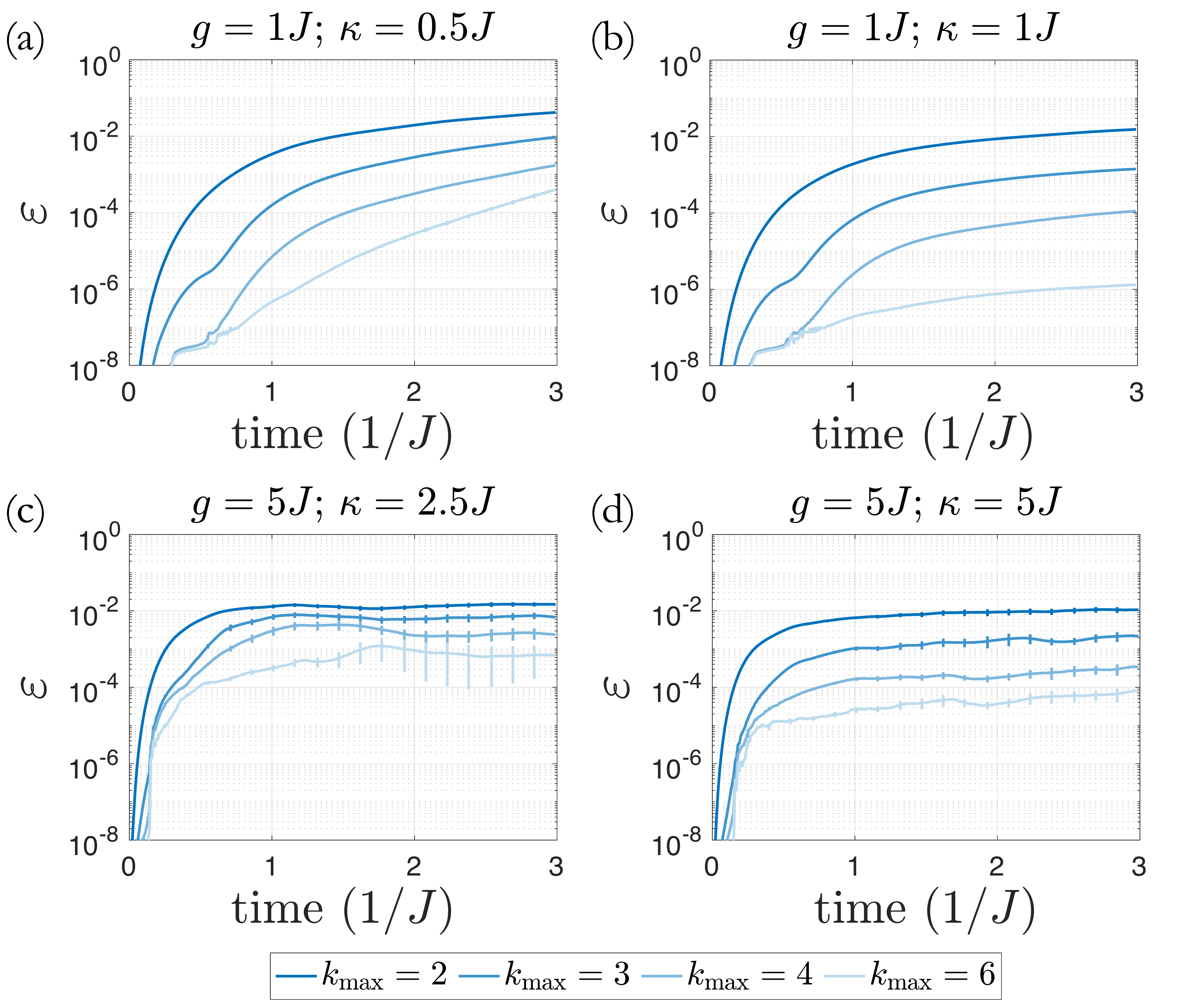}
\centering
\caption{Errors in the off-diagonal correlation functions in the Holstein model with $M=20$ lattice sites, upon limiting the maximum hierarchy depth of HOPS $k_{\rm max}$. Results taken with respect to $D=200$ and $k_{\rm max} = 8$. We have also incorporated conserved quantum numbers into the MPS algorithm~\cite{itensor}. These errors are averaged over 20 trajectories.}
\label{Fig_NBM_Kmax}
\end{figure}

The main numerical parameter for the HOPS algorithm is the depth of the hierarchy $k_{\rm max}$. Note that we use the same $k_{\rm max}$ for each hierarchy index in our many-body algorithm. We demonstrate the time-dependence of the errors in Fig.~\ref{Fig_NBM_Kmax}, where we can see that it is in fact possible to reach numerical precision while keeping the size of $k_{\rm max}$ relatively small, i.e., $k_\mathrm{max} <10$. We can see that for $g\sim J$ and smaller $\kappa$, i.e., slower decay of environment correlations and larger memory effects, that the errors grow rapidly, signifying the build up of non-Markovian effects. Intriguingly, for stronger coupling $g\sim5J $ we find the errors are strongly saturated indicating that perhaps strong coupling may actually be limiting the memory effects of the environment.

\begin{figure}[t!]
\includegraphics[width=8cm]{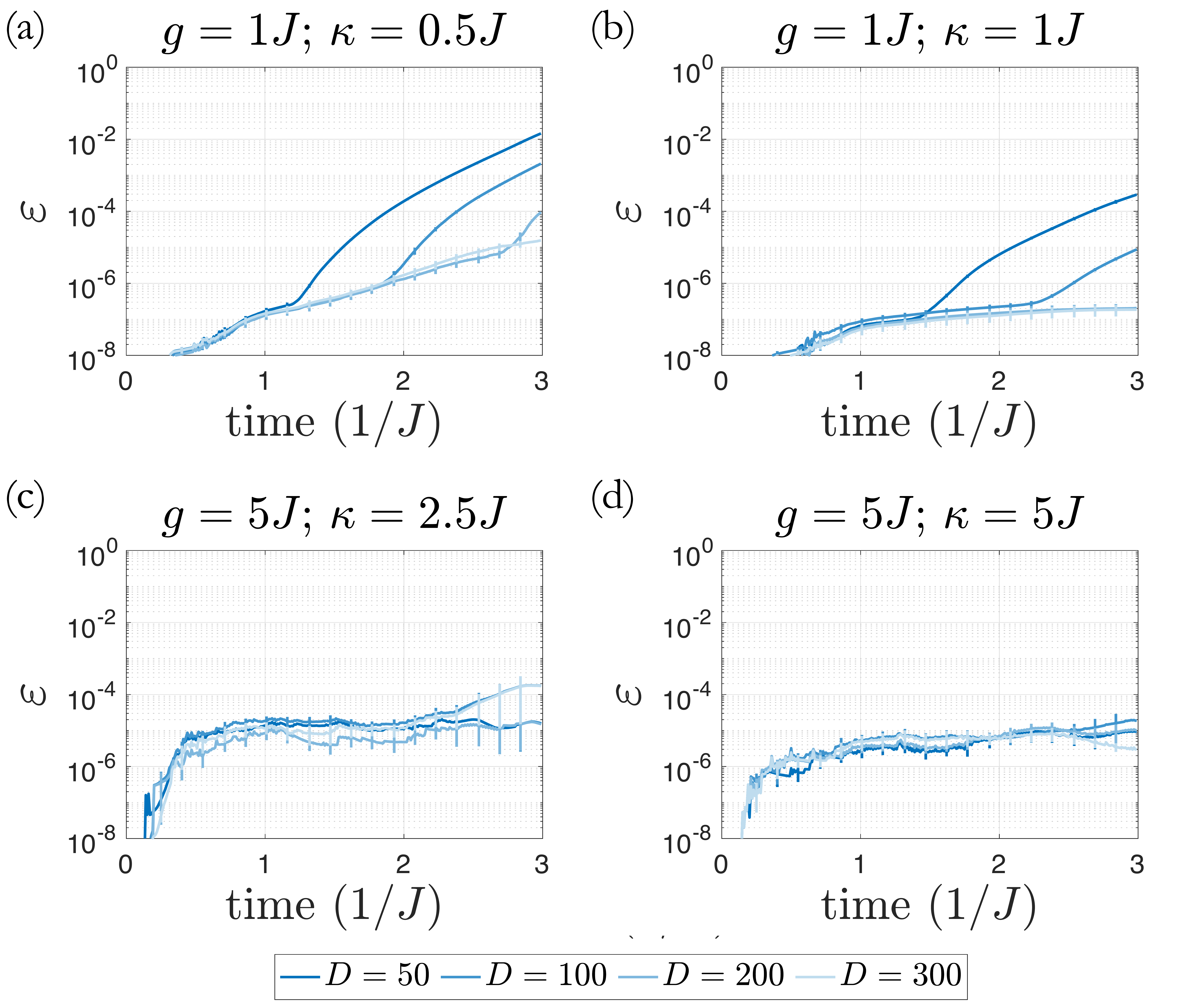}
\centering
\caption{Errors in the off-diagonal correlation functions in the Holstein model with $M=20$ lattice sites, upon limiting the MPS bond dimension $D$. Results taken with respect to $D=500$ and $k_{\rm max} = 8$. We have also incorporated conserved quantum numbers into the MPS algorithm~\cite{itensor}. These errors are averaged over 20 trajectories.}
\label{Fig_NBM_D}
\end{figure}

In our case we also have the additional consideration of the maximum allowed bond dimension of the MPS representation $D$. In Fig.~\ref{Fig_NBM_D} we plot the time-dependence of the errors (in the same observable) upon restricting this quantity, where again, we see that in principle we are able to obtain numerical precision for the short time dynamics after a global quench. For coupling strengths around $g\sim J $ we find similar behaviour as for closed system dynamics, where the errors grow approximately exponentially in time. Of course, this means that we are limited to short time dynamics after a global quench but this is also the case for conventional MPS simulations. In contrast, stronger coupling $g\sim5J $ we find a saturation of the errors, simply indicating that the strong coupling to the environment is suppressing the build up of correlations and entanglement in the dynamics. 

 \section{Numerical generation of colored noise}

In this section we describe one way of numerically generating the random noise terms that reproduces the desired statistics of the complex gaussian colored noise source $z^*(t) $ that we use in the HOPS algorithm. The derivation follows what is presented in Ref.~\cite{Gaspard:1999wx,PhysRevA.71.023812}

First we must define a response function $R(t)$,
\begin{equation}
R(t) = \frac{1}{2\pi} \int_{-\infty}^{\infty} d\omega G(\omega) e^{-i \omega t},
\end{equation}
where $G(\omega)$ is related to the correlation function through,
\begin{equation}
G(\omega) = \left( \int_{-\infty}^{\infty}dt \alpha(t) e^{i \omega t}\right)^{1/2}.
\end{equation}
Finally, we then calculate the colored noise $z(t) $ through,
\begin{equation}
z(t) = \int_{-\infty}^{\infty} ds R(s) \xi(t-s),
\end{equation}
where $\xi(\tau) = 1/\sqrt{2} \left( \xi'(\tau) + i \xi''(\tau) \right)$ and $\xi'(\tau)$ and $\xi''(\tau)$ are real independent gaussian white noise terms. The colored noise then has the desired properties,
\begin{equation}
\begin{split}
&\mathcal{E}\left[ z(t) z^*(t) \right] = \alpha(t - t'),\\
&\mathcal{E}\left[ z(t) z(t) \right] = 0.
\end{split}
\end{equation}

\section{Lindblad Master equation description}

In this section, 
we derive a simple Markovian master equation in the Lindblad form for the fermionic system only for an interaction Hamiltonian of the form considered in this letter,
\begin{equation}
\hat{H}_{\rm Int} = \sum_{n=1}^M \Big( \hat{L}_n\hat{a}^{\dagger}_n + \hat{L}_n^{\dagger}\hat{a}_n \Big).
\end{equation}

We first invoke the Born approximation and write the total density matrix of the whole system made up of the fermions and the phonons as $\rho_\mathrm{tot}(t) \approx \rho(t) \otimes \rho_E$, where $\rho(t)$ and $\rho_E$ are respectively the fermionic system and phonon environment density matrices~\cite{Breur_book}. At second order in the interaction Hamiltonian, the master equation for the system density matrix $\rho(t)$ reads in interaction picture~\cite{Breur_book,MilWise_Book}
\begin{equation} \label{Lindblad_SM}
 \partial_t \rho(t) = - \int^{t}_0dt' {\rm Tr_E} \Big( [ \hat{H}_{\rm Int}(t),[ \hat{H}_{\rm Int}(t'),\rho(t)\otimes \rho_E ] ] \Big),
\end{equation}
where ${\rm Tr_E}(\cdot)$ denotes the trace over the environment degrees of freedom and where,
\begin{equation}
\begin{split}
\hat{H}_{\rm Int}(t) & = e^{i(\hat{H}_s+\hat{H}_E)t}\hat{H}_{\rm Int}e^{-i(\hat{H}_s+\hat{H}_E)t} \\
& =  \sum_{n=1}^M \hat{L}_n(t) \hat{a}^{\dagger}_n(t) + \hat{L}_n^{\dagger}(t) \hat{a}_n(t),
\end{split}
\end{equation}
with $\hat{L}_n(t) = e^{i \hat{H}_s t} \hat{L}_n e^{-i \hat{H}_s t} $ and $\hat{a}_n(t) = e^{i \hat{H}_E t} \hat{a}_n e^{-i \hat{H}_E t}$.
Note that in Eq.~(\ref{Lindblad_SM}), we performed the Markov approximation, which consists in replacing $\rho(t')$ by $\rho(t)$ under the integral.

Expanding the interaction Hamiltonians explicitly in Eq.~(\ref{Lindblad_SM}) leads to
\begin{equation}
    \begin{aligned}
    \partial_t \rho(t)
    = -g^2 \int_0^t dt' & \Big\{\langle \hat{a}_n(t) \hat{a}^\dagger_n(t')\rangle \hat{L}^\dagger_n(t) \hat{L}_n(t') \rho(t) \Big. \\
    & + \langle \hat{a}_n^\dagger(t) \hat{a}_n(t')\rangle \hat{L}_n(t) \hat{L}_n^\dagger(t') \rho(t) \\
    & - \langle \hat{a}_n^\dagger(t') \hat{a}_n(t)\rangle \hat{L}_n^\dagger(t)  \rho(t) \hat{L}_n(t') \\
    & - \langle \hat{a}_n(t') \hat{a}_n^\dagger(t)\rangle \hat{L}_n(t)  \rho(t) \hat{L}_n^\dagger(t') \\
    & - \langle \hat{a}_n^\dagger(t) \hat{a}_n(t')\rangle \hat{L}_n^\dagger(t')  \rho(t) \hat{L}_n(t) \\
    & - \langle \hat{a}_n(t) \hat{a}_n^\dagger(t')\rangle \hat{L}_n(t')  \rho(t) \hat{L}_n^\dagger(t) \\
    &+ \langle \hat{a}_n(t') \hat{a}^\dagger_n(t)\rangle \hat{L}^\dagger_n(t') \hat{L}_n(t) \rho(t)  \\
    & \Big. + \langle \hat{a}_n^\dagger(t') \hat{a}_n(t)\rangle \hat{L}_n(t') \hat{L}_n^\dagger(t) \rho(t) \Big\}. \\
    \end{aligned}
\end{equation}
As we have a many-body system Hamiltonian $\hat{H}_s$, evaluating the term $\hat{L}_n(t) = e^{i \hat{H}_s t} \hat{L}_n e^{-i \hat{H}_s t} $ is not easy, as it requires to know its full energy spectrum. This is a general issue with applying open quantum systems to many-body Hamiltonians (but a major advantage of the HOPS algorithm considered in this work, as it does not require to calculate explicitly the system spectrum). Because of this, it is not possible to proceed further without having to introduce additional strong approximations on the behavior of $\hat{L}_n(t)$, if one wants to consider the environment correlation functions given in Eq.~(2) of the main text. For this reason, we assume here the limit of structureless phonon environments at thermal equilibrium, for which the correlation functions are given by
\begin{equation}
\begin{split}
\langle \hat{a}_n(t) \hat{a}_n^{\dagger}(t') \rangle &= (\bar{n} + 1)  \delta(t-t') \\ 
\langle \hat{a}_n^{\dagger}(t) \hat{a}_n(t') \rangle &= \bar{n} \delta(t-t'),
\end{split}
\end{equation}
where $\bar{n} = \left(\exp[\omega/(k_B T)] - 1 \right)^{-1}$ is the Bose factor with $k_B$ the Boltzmann constant and $T$ the temperature. This gives directly rise to the Lindblad master equation,
\begin{equation}
\begin{split}
 \partial_t \rho(t) &=  -i[\hat{H}_s,\rho(t) ] \\
 &+   g^2 (\bar{n} + 1 ) \sum_n \Big( \hat{L}_n \rho(t) \hat{L}^{\dagger}_n - \frac{1}{2}\left\{\hat{L}_n^{\dagger} \hat{L}_n , \rho(t)\right\} \Big) \\
 &+  g^2 \bar{n}  \sum_n \Big( \hat{L}_n^\dagger \rho(t) \hat{L}_n - \frac{1}{2}\left\{\hat{L}_n \hat{L}_n^\dagger , \rho(t)\right\} \Big),
 \end{split}
\end{equation}
where $\{ \hat{A} , \hat{B} \} = \hat{A}\hat{B} + \hat{B}\hat{A}$ denotes the anti-commutator. It is a quantum trajectory approach equivalent to this expression that we have used in the main text to compare to the results of HOPS.

\section{Environment correlations}

    Through features arising from the HOPS algorithm, it is also possible to calculate time-dependent observables within the environment~\cite{RevModPhys.89.015001}, thus enabling us to quantify the deviation away from the initial phonon population as well as capturing the build up of correlations between the system and the environment.  This can be seen using the starting point of the derivation of NMQSD, namely the expansion of the total state $|\Psi_T \rangle$ of the system and the environment in a basis of Bargmann coherent states of the environment~\cite{Suess:2015aa}. For the model considered in this paper of exponentially decaying correlation functions, one would normally model the environment of each site as a bath of harmonic oscillators with a Lorentzian spectral density and define the Bargmann coherent states according to this bath. Since here we work within the pseudo-mode picture of such a bath, we consider an expansion directly into a basis of pseudo-mode states $|z\rangle = |z_1, \cdots, z_M\rangle$ which reads (in interaction picture with respect to the bath)
\begin{equation}\label{expansion}
|\Psi_T \rangle = \int \frac{d^2z}{\pi} e^{-|z|^2} |\psi_z(t) \rangle \otimes |z\rangle,
\end{equation} 
with the shorthand $d^2z = d^2z_1 \cdots d^2z_M$ and $|z|^2 = \sum_n |z_n|^2$ and where $|\psi_z(t) \rangle = \langle z | \Psi_T \rangle $ is the system state whose the dynamics is governed by Eq.~(3) of the main text, where the subscript $z$ added here makes it clear that it is relative to a given environment state. The pseudo-mode states $|z_n\rangle$ are defined through $\hat{a}_n(t) |z_n\rangle = z_n(t) |z_n\rangle$, which define stochastic processes $z_n(t)$ whose ensemble average
\begin{equation}
    \mathcal{E}\left[ \dots \right] = \int \frac{d^2z}{\pi} e^{-|z|^2} \left[ \dots \right] 
\end{equation}
gives
\begin{equation}
    \mathcal{E}[ z_n(t) z_{n'}^*(t') ] = \langle \hat{a}_n^\dagger(t) \hat{a}_n'(t') \rangle = \delta_{n,n'} \alpha_n(t-t').
\end{equation}

Hence, any observable such as $\hat{a}^{\dagger}_n \hat{a}_n \hat{S}$ where $\hat{S}$ is an arbitrary system operator can be calculated via,
\begin{equation}
\begin{split}
\langle \hat{a}^{\dagger}_n &\hat{a}_n \hat{S} \rangle  = {\rm Tr}\left[   \hat{a}_n^{\dagger}(t)\hat{a}_n(t)  \hat{S} |\Psi_T \rangle \langle \Psi_T |   \right] \\
& = {\rm Tr}\left[   \hat{a}_n^{\dagger}(t)  \hat{S} |\Psi_T \rangle \langle \Psi_T | \hat{a}_n(t)  \right] - \langle \hat{S} \rangle \\
& = \int \frac{d^2z}{\pi} e^{-|z|^2} \Big( \langle \psi_z(t) | \otimes \langle z |  \hat{a}_n^{\dagger}(t)  \hat{S} |\Psi_T \rangle \langle \Psi_T |  \hat{a}_n(t)  \\
&~~~~~~~~~~~~~~| \psi_z(t) \rangle \otimes | z \rangle \Big) - \langle \hat{S} \rangle\\
&= \int \frac{d^2z}{\pi} e^{-|z|^2} \langle \psi_z(t) |\hat{S} |\psi_z(t) \rangle  |z_n(t)|^2 - \langle \hat{S} \rangle \\
&= \mathcal{E} \left[ \langle \psi_z(t) |\hat{S} |\psi_z(t) \rangle  |z_n(t)|^2  - \langle \hat{S} \rangle \right]
\end{split}
\end{equation} 
For $\hat{S}$ equal to the identity, this shows that the population of the pseudo-modes are related to the squared absolute values of the noise terms. One can thus track their time-dependent dynamics via monitoring of the noises used to generate the dynamical evolution of the system state. Any higher-order environment correlation functions can be obtained in a similar fashion. 
Note finally that since the non-linear version of HOPS has better convergence properties~\cite{Suess:2015aa, Hartmann:2017aa}, we used the expressions above with the system state of the non-linear HOPS Eq.~(6) of the main text with the transformed noises $\tilde{z}^*_n(t) = z^*_n(t) + \int_0^t ds \alpha^*_n(t - s) \langle \hat{L}_n^{\dagger} \rangle_s$ with $\langle \hat{L}^{\dagger}_n \rangle_s = \langle {\psi}^{(0)}(s) | \hat{L}^{\dagger}_n | {\psi}^{(0)}(s) \rangle$ instead of $z_n(t)$.

\end{document}